\documentclass[twocolumn, 
  aps, prl,
  amsmath,amssymb,
  ]{revtex4-1}
\usepackage{graphicx,color}
\usepackage{amsmath}
\usepackage{natbib}
\usepackage{epsfig}
\begin{document}

\title{\color{blue} Reduction of the Coulomb logarithm due to electron-neutral collisions}

\author{Sergey A. Khrapak}
\email{Sergey.Khrapak@dlr.de}
\affiliation{Institut f\"ur Materialphysik im Weltraum, Deutsches Zentrum f\"ur Luft- und Raumfahrt (DLR), 82234 We{\ss}ling, Germany; \\
Joint Institute for High Temperatures, Russian Academy of Sciences, 125412 Moscow, Russia}

\begin{abstract}
The frictional force (stopping power) acting on a test electron moving through the ideal electron gas is calculated taking into account electron-neutral atom collisions using the linear plasma response formalism. This allows us to elucidate how the effective Coulomb logarithm is affected by electron-neutral collisions. In agreement with a recent investigation by Hagelaar, Donko, and Dyatko [Phys. Rev. Lett. {\bf 123}, 025004 (2019)] we observe that the effective Coulomb logarithm decreases considerably due to electron-neutral collisions and becomes inversely proportional to the collision frequency in the highly collisional limit. 
%The exact numerical coefficient involved in this proportionality is obtained.    
\end{abstract}

\date{\today}

\maketitle

In a recent investigation Hagelaar, Donko, and Dyatko have shown that in partially ionized plasmas, Coulomb scattering can be significantly affected by electron-neutral collisions, and that this effect can be accounted for by a modification of the classical Coulomb logarithm~\cite{HagelaarPRL2019}. Frequent electron-neutral collisions result in a considerable reduction of the Coulomb logarithm. The proposed modification has been tested using first principles particle simulations, and the existing inconsistencies have been resolved when using the proposed modification.        

The suggested modification of the Coulomb logarithm was based on the detailed analysis of classical Coulomb collisions between two electrons with allowance for electron-neutral collisions (which scatter one of the electron isotropically) during a Coulomb collision event. An alternative approach to electronic transport, which can account for electron-neutral collisions self-consistently, is based on the linear plasma response formalism~\cite{NeufeldPR1955,ThompsonRMP1960}. For a recent example of its application see e.g. works on the problem of the ion drag force calculation in weakly ionized low-temperature dusty plasmas~\cite{IvlevPRL2004,IvlevPRE2005,FortovPR2005,KhrapakJAP2007}.      
In this context, it is useful to remind that the collision and the linear plasma response approaches are not competitive, but rather complementary, see e.g. Refs.~\cite{FortovPR2005,IvlevPPCF2004} for a detailed discussion. It is obviously desirable to examine the result of Ref.~\cite{HagelaarPRL2019} by an independent method.     

%In this way it does not account for the effect of electron-neutral collisions on the anisotropic component of the electric potential around    a moving electron~\cite{MontgomeryPlasmaPhys1968,StenfloPoF1973}, the anisotropy itself being completely neglected.

The purpose of this Rapid Communication is to complement the trajectory analysis of Ref.~\cite{HagelaarPRL2019} with a simple kinetic calculation using the linear plasma response formalism. We consider a simplest related problem. A test electron is moving through the electron gas immersed in an immobile neutralizing background (electron-electron collisions in the one-component plasma). In conventional weakly ionized plasmas electron transport can be significantly affected by electron-ion momentum transfer, but this was not included in the analysis of Ref.~\cite{HagelaarPRL2019} and is omitted for consistency here. Neutral atoms are also present in the system under consideration and the electron-neutral collision frequency is $\nu$. This frequency $\nu$ can vary considerably. We calculate the frictional force (stopping power) acting on such a test electron starting from collisionless limit ($\nu\rightarrow 0$) and then follow how this force changes when $\nu$ increases. In this way an effective Coulomb logarithm can be introduced and its dependence on the electron-neutral collision frequency can be determined. The modification of the Coulomb logarithm in the highly collisional regime is consistent with that proposed in Ref.~\cite{HagelaarPRL2019}, except a small difference in the numerical coefficient involved. 

The starting point of our simple calculation is the solution for the electrostatic potential around an immobile electron immersed in a system of charged particles 
\begin{equation}\label{potential}
\phi({\bf r})=-\frac{e}{2\pi^2}\int d{\bf k}\frac{e^{i{\bf k} \cdot {\bf r}}}{k^2\epsilon(k,\omega)|_{\omega=0}},
\end{equation}       
where $-e$ is the electron charge, ${\bf k}$ is the wave vector, $\omega$ is the frequency, and $\epsilon(k,\omega)$ is the system permittivity. In the following an ideal electron gas is considered so that electron-electron correlations can be completely neglected. The condition $\omega=0$ indicates that the test charge is at rest. Electron-neutral collisions will have no effect on the distribution of electrostatic potential around an immobile electron, when the surrounding plasma is at rest and isotropic. Assume now that the test electron is moving with a velocity ${\bf u}$ through the electron gas (its energy is not too high, so that the classical consideration is appropriate). The problem is equivalent to the electron at rest immersed in an electron gas moving with a velocity $-{\bf u}$ relative to it. Equation~(\ref{potential}) still applies, but with a condition $\omega = -{\bf k}\cdot {\bf u}$. The frictional force (stopping power) is the force that the test electron experiences in its own induced field
\begin{equation}
F_{\rm st}=e \frac{\partial \phi}{\partial z}|_{{\bf r}=0},
\end{equation}
where $z$-axis is parallel to the direction of electron motion.     
After simple algebra we arrive at the general expression for the force
\begin{equation}\label{force}
|F_{\rm st}|=\frac{e^2}{\pi}\int_0^{k_{\max}}\int_{-1}^{1}kg {\rm Im}[\epsilon(k,-{\bf k}\cdot {\bf u})^{-1}]dkdg,
\end{equation}
where $g=\cos\theta$, and $\theta$ is the angle between ${\bf k}$ and ${\bf u}$. The frictional (stopping) force acts in the direction opposite to ${\bf u}$. 

We employ a simplest model for the system permittivity, which is appropriate for superthermal test electron (when velocity distribution is unimportant). The plasmon permittivity accounting for the effect of electron-neutral collisions reads 
\begin{equation}\label{plasmon}
\epsilon(k,\omega)= 1-\frac{\omega_{\rm p}^2}{\omega(\omega+i\nu)},
\end{equation}
where $\omega_{\rm p}=\sqrt{4\pi e^2 n/m}$ is the electron plasma frequency, $n$ is the electron density, and $m$ is the electron mass. Following Ref.~\cite{HagelaarPRL2019}, the electron-neutral collision frequency $\nu$ is assumed constant.  Eq.~(\ref{plasmon}) follows straightforwardly from the linear perturbation analysis of fluid continuity and momentum equations coupled to the Poisson equation. Thus, it is clearly relevant in the collisional regime. Its relevance to the collisionless regime is, however, less evident and will be discussed separately. The effect of electron deceleration on the frictional force~\cite{KompaneetsPoP2014} is neglected. 

The $k$-integration in Eq.~(\ref{force}) is cut off at the value $k_{\rm max}$, corresponding to short separations between colliding electrons, which cannot be treated within the plasma response formalism. It is generally assumed that $k_{\rm max}\simeq 1/\rho_0=\mu u^2/e^2$, where $\mu=m/2$ is the reduced mass for electron-electron collisions, and $\rho_0$ is the characteristic Coulomb (Landau) length (distance at which the kinetic energy of colliding electrons becomes comparable to the Coulomb interaction energy). 

There are the other two important inverse length scales of interest for the problem at hand, which are suggested by the form of the permittivity considered. 
The first is the inverse mean free path of the test electron with respect to collisions with neutrals, $k_{\rm coll}=\nu/u$. For $k>k_{\rm coll}$ scattering occurs in the collisionless regime, for $k<k_{\rm coll}$ collisions are important. The second is the minimum wave-number $k_{\rm min}=\omega_{\rm p}/u$, corresponding to collisionless contribution to the stopping force. 
The strong inequality $k_{\rm min}\ll k_{\rm max}$ is satisfied in the ideal electron gas regime. The exact origin and the physical meaning of $k_{\rm min}$ will become apparent shortly.     
 
\begin{figure}
\includegraphics[width=8.cm]{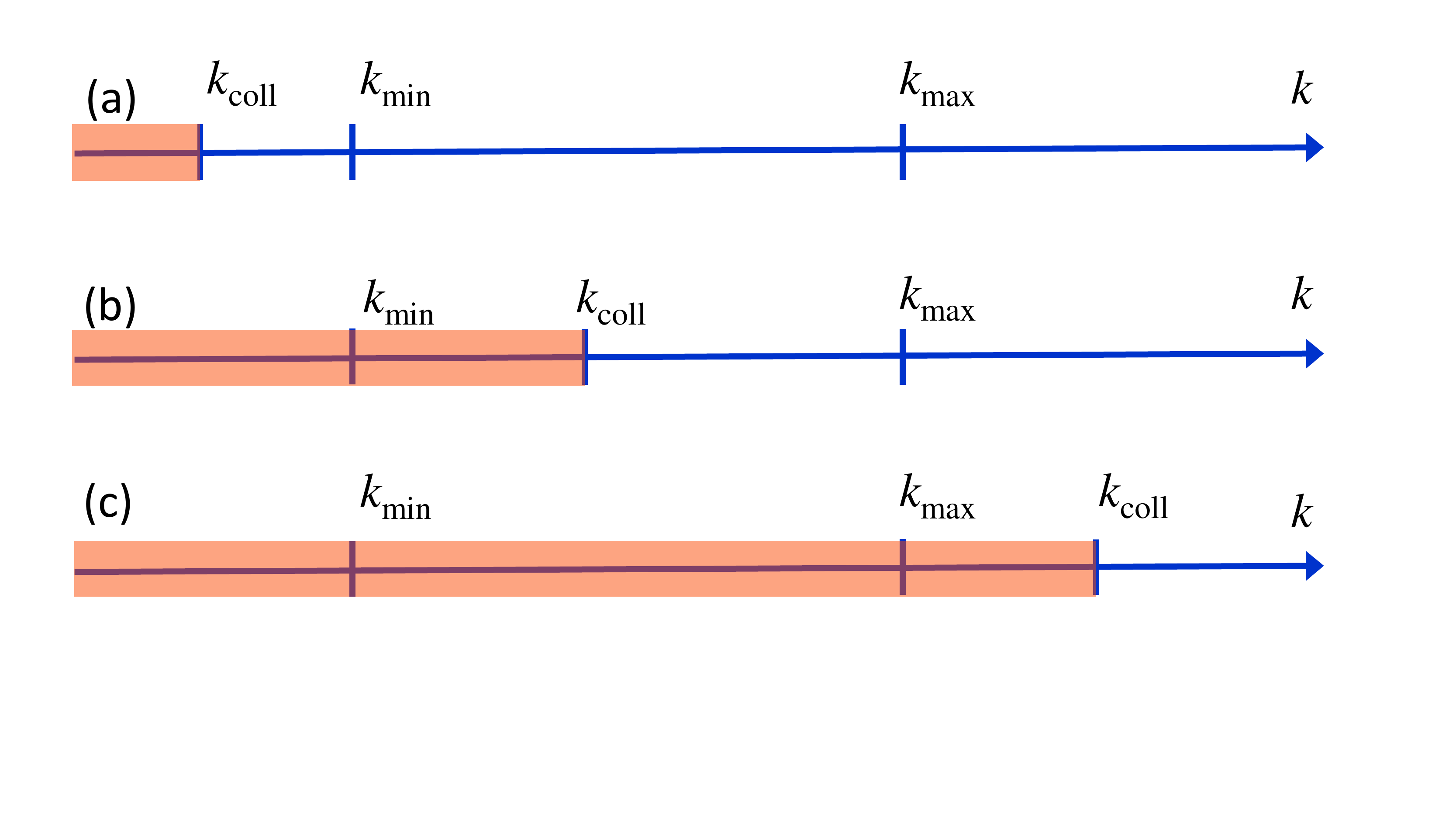}
\caption{Three different regimes considered in terms of the relationship between  $k_{\rm coll}$, $k_{\rm min}$, and $k_{\rm max}$. Regime (a) corresponds to weak collisionality, $k_{\rm coll}<k_{\rm min}\ll k_{\max}$; regime (b) corresponds to intermediate collisionality, $k_{\rm min}< k_{\rm coll}< k_{\rm max}$; and (c) is the highly collisional regime with $k_{\rm coll} > k_{\rm max}$. The collisional regime is highlighted by color.}
\label{Fig1}
\end{figure}

In the following three regimes will be considered. These differ by the relationship between $k_{\rm coll}$, $k_{\rm min}$, and $k_{\rm max}$, as sketched in Fig.~\ref{Fig1}. 

(i) {\it Collisionless regime}. This regime corresponds to the condition $k_{\rm coll}<k_{\rm min}\ll k_{\rm max}$ (see Fig.~\ref{Fig1}a). Substitution $\omega=-{\bf k}\cdot {\bf u}=-kug$ in Eq.~(\ref{plasmon}) gives
\begin{displaymath}
\epsilon(k,-kug)\simeq 1-\frac{\omega_{\rm p}^2}{(kug)^2}-i\frac{\nu \omega_{\rm p}^2}{(kug)^3} = 1-\frac{\omega_{\rm p}^2}{(kug)^2}-i\delta.
\end{displaymath} 
The real part corresponds to the dispersion relation of the propagating plasmon colective mode, the small imaginary part describes weak damping of this mode. The main contribution to the integration in Eq.~(\ref{force}) will come from the pole given by the dispersion relation $\epsilon(k,-{\bf k}\cdot {\bf u})=0$, which corresponds to the interaction via plasma waves~\cite{LL}. We can use the relation
\begin{displaymath}
{\rm Im}\lim_{\delta\rightarrow 0}\frac{1}{z-i\delta}=\lim_{\delta\rightarrow 0}\frac{\delta}{z^2+\delta^2}=\pi \delta(z)
\end{displaymath}   
to obtain
\begin{equation}\label{collless}
F_{\rm st}={e^2}\int_0^{k_{\max}}\int_{-1}^{1}kg \delta\left(1-\frac{\omega_{\rm p}^2}{k^2u^2g^2}\right)dkdg.
\end{equation}
Next, we use the property
\begin{displaymath}
\delta\left(1-\frac{\alpha^2}{x^2}\right)=\frac{\alpha}{2}\left[\delta(x-\alpha)+\delta(x+\alpha)\right]
\end{displaymath} 
and take into account that because $|g|<1$, the $k$ integration in Eq.~(\ref{collless}) should start from $k_{\min}=\omega_{\rm p}/u$. This yields
\begin{equation}\label{collless1}
F_{\rm st}=\frac{e^2}{\omega_{\rm p}^2 u^2}\int_{k_{\min}}^{k_{\max}}\frac{dk}{k}=\frac{4\pi e^4 n}{mu^2}\ln\left(\frac{u}{\omega_{\rm p}\rho_0 }\right).
\end{equation} 
The classical Coulomb logarithm $\ln\left(\lambda_{\rm D}/\rho_0\right)$ emerges if we make the substitution $T\rightarrow mu^2$ in the expression for the electron Debye radius, $\lambda_{\rm D}=\sqrt{T/4\pi e^2 n}$. This is a known property of screening of a fast projectile in a plasma~\cite{KhrapakPoP2005,LudwigNJP2012}. The Coulomb logarithm is large since the ideal electron gas is considered (which ensures that $\lambda_{\rm D}\gg\rho_0$). We observe that electron-neutral collisions do not contribute to the stopping force as long as $k_{\rm coll}<k_{\rm min}$, that is when $\nu<\omega_{\rm p}$.  

An important observation is that expression (\ref{collless1}) is independent of the concrete form of the imaginary contribution to $\epsilon(k,\omega)$, provided it is small enough. The main contribution comes from the pole given by  $\epsilon(k,\omega)=0$, which corresponds to electron-electron interaction via the plasmon collective mode. The result, identical to Eq.~(\ref{collless1}), could be obtained with the plasma permittivity of completely collisionless plasma, expressed via the conventional plasma dispersion function~\cite{PeterPRA1991}. In this case the imaginary part comes from the (exponentially small) Landau damping term. Note that in Ref.~\cite{PeterPRA1991} the contributions from the collective regime $k\lambda_{\rm D}<1$ and individual regime $k\lambda_{\rm D}>1$ are treated separately. Their sum coincides with Eq.~(\ref{collless1}), and this result is often quoted as Bohr stopping force~\cite{PeterPRA1991}. 
  
(ii) {\it Highly collisional regime}. In this regime electrons are highly collisional in the entire range of $k$, since $k_{\rm coll}>k_{\rm max}$ (see Fig.~\ref{Fig1}c). Eq.~(\ref{plasmon}) gives
\begin{displaymath}
\epsilon(k,-kug)\simeq 1-\frac{i \omega_{\rm p}^2}{\nu kug}.
\end{displaymath}    
Substituting this into Eq.~(\ref{force}) we get
\begin{equation}\label{collisional1}
F_{\rm st}=\frac{e^2\omega_{\rm p}^2u\nu}{\pi}\int_0^{k_{\max}}\int_{-1}^{1}\frac{k^2g^2 dk dg}{(\nu kug)^2+\omega_{\rm p}^4}.
\end{equation}
The integration over $g$ then yields
\begin{equation}\label{collisional2}
F_{\rm st}=\frac{2e^2\omega_{\rm p}^4}{\pi(u\nu)^2}\int_0^{x_{\max}}dx\left[1-\frac{1}{x}{\rm ArcTan}(x)\right],
\end{equation}
where $x= ku\nu/\omega_{\rm p}^2$, $x_{\rm max}=u_0(\lambda_{\rm D}/\rho_0)(\nu/\omega_{\rm p})$, $u_0=u/v_{\rm T}$, and $v_{\rm T}=\sqrt{T/m}$. The considered regime corresponds to $u_0>1$, $\lambda_{\rm D}\gg \rho_0$, and $\nu> \omega \sim \omega_{\rm p}$. Thus, we must assume $x_{\max}\gg 1$. In this case the integral can be asymptotically expanded as
\begin{displaymath}
\int_0^{x_{\max}}dx\left[1-\frac{1}{x}{\rm ArcTan}(x)\right]\simeq x_{\max}-\frac{\pi}{2}\ln x_{\max}-\frac{1}{x_{\max}}+...
\end{displaymath}  
Keeping the first dominant term we finally get
\begin{equation}\label{collisional3}
F_{\rm st} = \frac{8e^4n}{mu^2}\frac{u}{\nu\rho_0}
\end{equation}  
in the considered highly collisional regime. 

(iii) {\it Intermediate collisionality}. Taking into account the results already obtained, the friction force can be estimated as a sum from collisional and collisionless contributions, with appropriate integration limits (see Fig.~\ref{Fig1}b). The result is 
\begin{equation}\label{intermediate}
F_{\rm st} = \frac{4\pi e^4 n}{mu^2}\ln\left(\frac{u}{\nu\rho_0}\right)+\frac{8e^4n}{mu^2}.
\end{equation}
The first term corresponds to collisionless contribution with the collisional modification of the low-$k$ integration limit. Collisions diminish the argument of the Coulomb logarithm from $u/\omega_{\rm p}\rho_0\equiv k_{\rm max}/k_{\rm min}$ to $u/\nu\rho_0\equiv k_{\rm max}/k_{\rm coll}$. The second terms is the contribution from the collisional domain with $x_{\rm max}=k_{\rm coll}u\nu/\omega_{\rm p}^2=\nu^2/\omega_{\rm p}^2$. 

We now have a complete picture of how the effect of electron-neutral collisions affects the magnitude of the stopping power and, hence, the effective Coulomb logarithm. When $\nu<\omega_{\rm p}$ (weak collisionality) the effect of collisions is negligibly small. The Coulomb logarithm is large. At $\omega_{\rm p}<\nu< u/\rho_0$ (moderate collisionality) collisions reduce the magnitude of stopping power, but the Coulomb logarithm remains large, approaching unity at $\nu\sim u/\rho_0$. At $\nu > u/\rho_0$ (high collisionality), the effective Coulomb logarithm becomes inversely proportional to the collision frequency and drops below unity. The sequence of modifications it experiences looks approximately as
\begin{displaymath}
\ln\left(\frac{u}{\omega_{\rm p}\rho_0}\right)\rightarrow \ln\left(\frac{u}{\nu\rho_0}\right)+\frac{2}{\pi} \rightarrow \frac{u}{\nu\rho_0}.
\end{displaymath}
This is similar to what have been proposed in Ref.~\cite{HagelaarPRL2019} on the basis of trajectory analysis. From the binary collision perspective, the friction force in the collisionless regime can be simply estimated as the product of the electron flux $nu$, momentum carried by each electron $\mu u$ (in the center of mass reference frame), and the Coulomb scattering cross section $4\pi\rho_0^2\ln\Lambda$:
\begin{equation}\label{bc}
F_{\rm bc}= \frac{8\pi e^4 n}{mu^2}\ln\Lambda.
\end{equation}           
Comparing Eqs.~(\ref{collisional3}) and (\ref{bc}) we obtain the effective Coulomb logarithm in the highly collisional regime as
\begin{equation}
\ln\Lambda_{\rm eff}=\frac{1}{\pi}\frac{u}{\nu\rho_0}.
\end{equation}
This coincides with Eq.~(15) from Ref.~\cite{HagelaarPRL2019}, except one minor but interesting detail: the numerical constant $1/3$ suggested in \cite{HagelaarPRL2019} is evaluated here as $1/\pi$ using the linear plasma response formalism.       

To conclude, the modification of the Coulomb logarithm due to electron-neutral collisions in partially ionized plasma proposed in Ref.~\cite{HagelaarPRL2019} has been scrutinized using the linear plasma response formalism. The presented alternative calculation demonstrates full agreement with the results from ~\citep{HagelaarPRL2019}, except a minor difference in the numerical coefficient. Close agreement between the two independent theoretical approaches and successful tests against the first principles particles simulations provide strong confidence regarding the important effect of electron-neutral collisions on the electron-electron collision term discussed here. The underlying physics is, however, quite general and similar mechanisms and scenarios can operate for other plasma-related problems. This can constitute important direction for future work. In particular, experimental verification of the effect discussed would be highly desirable.   
  
%\acknowledgments
I would like to thank Alexei Ivlev for useful suggestions and Victoriya Yaroshenko for a careful reading of the manuscript.

\bibliographystyle{aipnum4-1}

\bibliography{CL_References}

\end{document}